\documentstyle[manuscript,aps,prb]{revtex}
\tighten
\begin{document}
\title{Electronic structure of polychiral carbon nanotubes}
\author{Ph. Lambin$^a$, V. Meunier$^b$, and A. Rubio$^c$}
\address{$^a$ D\'epartement de physique, FUNDP, 61 Rue de Bruxelles, 
B 5000 Namur, Belgium.\\
$^b$ Department of Physics, North Carolina State University, 
Rayleigh NC 27695, USA.\\
$^c$ Departamento F\'\i sica Te\'orica, Universidad de Valladolid,
E 47011 Valladolid, Spain\\ and Donostia International Physics
Center, San Sebastian, Spain}
\date{\today}
\maketitle

\begin{abstract}
Most of the works devoted so far to the electronic band structure of multiwall
nanotubes have been restricted to the case where the individual layers have
the same helicity. By comparison, much less is known on the electronic
properties of multiwall nanotubes that mix different helicities. These are
interesting systems, however, since they can be composed of both metallic and
semiconducting layers. For the present work, tight-binding calculations were
undertaken for polychiral two-layer nanotubes such as (9,6)@(15,10),
(6,6)@(18,2), and others. The recursion technique was used to investigate how
the densities of states of the individual layers are affected by the intertube
coupling. Constant-current STM images were also calculated for these
systems. The result obtained is that the image of a two-wall nanotube is
pretty much the same as the one of the isolated external layer. It is only in
the case of monochiral, commensurate structures like (5,5)@(10,10) that
interlayer effects can be seen on the STM topography.
\end{abstract}

\section{Introduction}
Multiwall nanotubes (MWNT) have not attracted as much attention from the
theoreticians as the single-wall carbon tubes did, although they may be useful
in many applications. Multiwall nanotubes are obviously more complex than the
one-layer tubules, which explains why the former are much less documented. As
far as the electronic structure is concerned, calculations have already been
performed for the simplest MWNTs, those made of non-chiral layers
.\cite{Saito93,Charlier93,My113,My126,Ostling97,Kwon98,Rub99} The systems that
were investigated in these works were all made of either zig-zag or armchair
nanotubes, without mixing. The reason was simply that mixing nanotubes with
incommensurate periods along their axis leads to an non-periodic system, which
therefore precludes the use of Bloch theorem and makes the calculations more
difficult. However, there are many indications from electron
diffraction~\cite{Iijima91,Zhang93,Liu94,Iijima94} and
STM~\cite{Ge93,Hassanien99} that the layers in a multiwall nanotube
often have different helicities with nearly random distribution.

The aim of the present paper is to investigate the electronic properties of
polychiral nanotubes, namely multiwall structures mixing layers with different
helicities. The motivation of this work is twofold. First, to find out in
which way electronic states can be induced in the band gap of a semiconducting
layer by its coupling to a metallic tube of different chirality. The resulting
interlayer coupling varies from site to site in a pseudo-random manner that
could be responsible for localization of the electronic wave
functions.\cite{Anderson58,Lee85} No such effects were found in the present
study. The second motivation was to see whether the electronic local density
of states of the external layer of a MWNT can reflect the atomic structure of
the underlying layers, leading to a pattern that could be observed with an
STM. In some cases, MWNT analyzed with the STM appear like graphite, where
only every other two atoms of the external layer are clearly
imaged.\cite{Lin96,Biro98} In other cases, there is a spatial modulation of
the image intensity, most obviously because a perfect lattice coherence cannot
be realized between two cylindrical graphitic sheets when the layers have
different helicities.\cite{Ge93} Our calculations show that, indeed, a site
asymmetry of the STM image of a MWNT similar to that of graphite may appear,
but this requires a special monochiral geometry like (5,5)@(10,10). In
polychiral nanotubes, by contrast, there is no site asymmetry and no Moir\'e
pattern in the STM images computed for bilayer systems. This conclusion is in
agreement with recent STM images with atomic resolution obtained on
MWNTs,\cite{Hassanien99} although Moir\'e patterns have frequently been
observed in other experiments as recalled here above.

All these effects were explored within a tight-binding description of the
$\pi$ electronic states, using the methodology presented in section 2. The
results on the local DOS calculations for nanotubes mixing semiconducting and
metallic layers are given in sections 3-4, and the STM image simulations are
reported to section 5. All the calculations were restricted to bilayer
nanotubes to keep the computing load reasonable.

\section{Methodology}\label{meth}
Several two-wall nanotubes were generated on the computer, with the
requirement that the layer radii differ by approximately 0.34 nm which
corresponds to the observed interlayer distance in MWNTs. In the
$\pi$-electron tight-binding Hamiltonian used, the first-neighbor C pairs
within a same shell received a hopping interaction $\gamma_0$ = -2.75 eV. This
value was used for consistency with previous calculations of ours. That value
slightly underestimates a recent experimental estimation of $\gamma_0$ (-2.9
eV),\cite{Saito00} which simply means that our energies should be scaled by a
factor 1.05. This scaling does not alter any of the conclusions of the present
work related to the interlayer interaction and STM imaging.  The interlayer
interactions were written as $W \cos\phi \exp{[-(d-\delta)/L]}$ with $d$ the
distance between the coupled atoms, $\phi$ the angle between the $\pi$
orbitals on these two atoms, $\delta$ = 0.334~nm, and $L$ = 0.045~nm. Two
values of $W$ where used to describe the graphite AA, BB or AB-like
interactions: $W$ = 0.36~eV for the first two and $W$ = 0.16~eV for the
latter.\cite{My126} The range of the interlayer interactions was limited to a
maximum distance $d$ = 0.39~nm. This parameters reproduce well
first-principles calculations for MWNTs.\cite{Charlier93}

Local densities of states in the multiwall nanotubes were computed by the
recursion method.  This technique does not rely on the Bloch theorem. It gives
rise to a continued-fraction development of the Green's function diagonal
elements in the complex energy plane, $G_{ii}(z) = \langle i | (z-H)^{-1} |i
\rangle$. For each atomic site $i$ of interest, $n$ = 500 levels of continued
fraction were computed. When the continued fraction is truncated after $n$
levels, the resulting density of states is composed of $n$ Dirac delta
peaks. With $n-1$ levels, another set of $n-1$ peaks is obtained. These two,
interlaced sets were mixed with equal weights, and each delta peak was
represented by a Gaussian function with standard deviation 0.023 eV (the band
width, $6|\gamma_0|$, divided by $\sqrt{2}n$). Due to this broadening, all the
singularities of the densities of states, including the band edges, are
slightly smoothed out.  Due to its smallness, this broadening should not alter
a main conclusion of the work, namely the absence of localized states in the
gap of the semiconducting layer.

The change of density of states brought about by the interlayer coupling is
expected to be small. The first-order perturbation expression of the Green
function is indeed $G = G_0 + G_0W G_0$. The unperturbed Green function
$G_0$ is made of blocks corresponding to the individual layers that $W$
couples together. Due to that structure, all the diagonal elements of
$G_0W G_0$ are zero, which means that the density of states is not
perturbed at first order in the interlayer interaction.

STM image simulations of multiwall nanotubes were performed for comparison with
experiment. These calculations are based on a simple tight-binding theory of
the STM current~\cite{My174}
\begin{displaymath}
I = (2\pi)^2 \; \frac{e}{h} \int_{E_F^s-eV}^{E_F^s} dE \; n_t(E_F^t-E_F^s+eV+E)
 \sum_{i,i'\in s} v_{ti} v_{ti'}^*n_{ii'}^s(E)
\end{displaymath}
where the $E_F$'s are the Fermi levels of the tip (t) and sample (s) and $V$
is the tip--sample bias potential. The tip is treated as a single atom with an
s orbital and a Gaussian density of states $n_t(E)$. $v_{ti}$ is the
tight-binding hopping interaction between the tip atom and $\pi$ orbital
located on site $i$ of the nanotube sample, and $n_{ii'}^s(E) = (-1/\pi)
\mbox{ Im } G_{ii'}^s(E)$. The Green function elements $G_{ii'}$ of the
nanotube were computed by recursion with 200 continued-fraction levels that
give converged results for the present imaging studies.  Similarly to the
density of states calculations, a small imaginary part was added to the
energy to force the convergence.

\section{Metal-semiconductor nanotubes} \label{case1}

We first consider two-wall nanotubes having a metal at the inner shell and a
semiconductor at the outer shell.  Table~\ref{list1} gives a few such
metal-semiconductor nanotubes. The first system mixes an armchair and a
zig-zag nanotube. Although these two nanotubes are non-chiral, their chiral
angles differ by 30$^{\circ}$, their translation periods differ by a factor of
$\sqrt{3}$, and the combined system may be described as polychiral. The next
three nanotubes of table~\ref{list1} are real polychiral systems. By contrast,
the fifth nanotube is monochiral since its layers have the same
helicity. Hence, the two layers in the (15,-6)@(15,10) nanotube have the same
Bravais period but have opposite chiral angles. This last nanotube differs
from the fourth one by the fact the inner layers (15,-6) and (9,6) are
enantiomers.

Local density of states in the external layer were computed in a slice of 0.2
nm height, which contained between 32 and 40 atoms, depending on the
nanotube. Although the coupling to the inner layer varies from site to site,
all the atoms of the external layer were found to have pretty much the same
density of states. 

For all the polychiral nanotubes investigated, the density of states of the
semiconducting layer was found to be weakly affected by its coupling to the
inner metallic layer, at least in an interval between -1 and +1 eV (the zero
of energy is always considered to be at the Fermi level). In that interval of
energy, the metallic tube presents a constant density of states -- hereafter
called the metallic plateau -- with no van Hove
singularities. Fig.~\ref{tub4/5}(a), which concerns with (15,-6)@(15,10), is a
typical example of this effect.  As compared to the single-wall (15,10)
nanotube (dashed curve), there is a minute downshift of the bottom of the
conduction band of the semiconducting layer (full curve), whereas the top of
the valence band does not move. The shapes of the gap edges are not
modified. If the hopping interactions between the layers introduce a tailing
of the valence and conduction states inside the gap, the decay will take place
in an energy range shorter than the peak broadening used (0.023 eV). The
metallic layer induces a few states in the band gap of the semiconductor (the
total number of states in the band gap is 0.42$\times10^{-4}$ per atom of the
semiconducting layer, see table~\ref{list1}). The density of states in the gap
region is approximately uniform and very small.

The interlayer coupling is much more efficient in the monochiral (9,6)@(15,10)
nanotube, as shown in fig.~\ref{tub4/5}(b). The density of states in the band
gap region is approximately five times larger than in fig.~\ref{tub4/5}(a) and
reaches 2.5\% of that of the metallic layer. It must be stressed out that both
nanotubes in fig.~\ref{tub4/5} exhibit exactly the same electronic structure
when the interlayer coupling is switched off. In other words, all the
differences between the full curves in figs.~\ref{tub4/5}(a) and (b) come out
from the different environments the (15,10) layer feel in both systems. The
monochiral tubes constitute a particular case in which the electronic
properties are much affected by specific symmetries in MWNTs, both as
pseudogaps in the local density of states or as a change in the intensity of
every two atom of the STM image (see below). A change of the width of the
metallic gap was also found in the case of commensurate three-wall armchair
tubes.\cite{My185}

\section{Semiconductor-metal nanotubes} \label{case2}

We now consider two-wall nanotubes having the semiconducting layer at the
interior and the metallic layer outside. A list of such nanotubes which mix
different helicities is given in table~\ref{list2}. Local density of states in
the inner layer were computed in a slice of 0.3 nm height, which contained
between 24 and 32 atoms depending on the nanotube. Here again, the
fluctuations of densities of states in the semiconducting layer, although two
times as large as in Sect.~\ref{case1}, remained small (see
table~\ref{list2}). The local densities of states were then averaged as
before.

The ratio between the band gap of the semiconductor and the width of the
metallic plateau of the metal is now approximately 2:3, instead of 1:6 as for
the previous configurations (Sect.~\ref{case1}). For instance, the metallic
plateau of the (10,10) nanotube is bounded by two Van Hove singularities at $E
= \pm$0.9~eV. These singularities can be seen in the density of states of the
(6,4) layer in the (6,4)@(10,10) bilayer (fig.~\ref{tub6}(a)). In the
conduction band for instance, the singularity leads to a resonance followed by
an anti-resonance. This kind of structure was frequently observed among the
nanotubes of table~\ref{list2}. In a systematic way, also, the band gap of the
semiconductor is reduced by the coupling to the outer layer: the top of the
valence band has moved upwards by approximately 0.03 eV.

As shown in fig.~\ref{tub6}(a), the density of states in the band gap
of the inner semiconducting tube is small, but still 
significantly higher than in fig.~\ref{tub4/5}(a) for
instance. The difference between these two situations is that the
semiconducting layer is now at the interior rather than at the exterior, and
the inner layer is more perturbed than the outer one. Indeed the average
numbers of intersheet bonds per atom in layers 1 and 2 are inversely
proportional to the number of atoms in these layers. Since there are
approximately two times less atoms in the inner shell than in the outer (for
those systems we are investigating), the interlayer coupling is two times more
efficient on layer 1 than on layer 2. Since, in addition, the (6,4)
semiconductor has a larger band gap than those of the semiconducting layers of
table~\ref{list1} (due to its smaller diameter), the number of states in the
gap has increased. As revealed by table~\ref{list2}, the number of states in
the gap looks remarkably constant for all the nanotubes examined, around
2.3$\times10^{-4}$ per atom.

As for the metallic layer, one can hardly see any change in its density of
states around the Fermi level (fig.~\ref{tub6}(b)). The site-to-site
fluctuations of densities of states near the Fermi level are less than 1\% of
the (10,10) density of states at $E_F$. This kind of weak disorder, here due
to the coupling with a chiral nanotube, is small but is perhaps sufficient to
affect the transport properties of a MWNT in the weak localization
regime.\cite{Langer96,Schonenberger99} Clear effects of the coupling to the
inner layer appear below -0.5~eV and above +0.5~eV, where the (6,4) nanotube
has its band edges. It is clear from this example that the amplitudes of the
Van Hove singularities at $E = \pm$0.9~eV are reduced as compared with the
single-wall nanotube. This is due to the breaking of the translational
symmetry brought about by the coupling between layers of different
chiralities.

\section{Simulation of the STM image of two-wall nanotubes}\label{STMsec}

Several constant-current STM images of multiwall nanotubes show intensity or
contrast modulations.\cite{Ge93,Biro98} Such modulations have been interpreted
as being a Moir\'e pattern formed by the atomic structure of the last
two layers. In graphite, Moir\'e pattern effects have clearly been identified
with an STM in regions where the last layer was folded back on the surface
with a misorientation of its crystallographic directions.\cite{Beyer99}

As pointed out in the previous sections, the local DOS show only little
variations on going from one site to another in a multiwall nanotube. To
quantify that property, the fluctuations (rms) of the local DOS were computed
on a chain of 25 first-neighbor atoms located as close as possible to a
generator of the external layer. These atoms were selected because they would
be probed in a scan of the topmost part of the nanotube by a STM tip. The DOS
fluctuations, averaged over the energy interval (-1,+1) eV, are listed in the
last row of tables~\ref{list1} and \ref{list2}. They are small, less than 1\%
of the mean density of states. According to these data, the spatial variations
of the density of states in polychiral nanotubes cannot explain the
modulations of the STM image intensity.

Fig.~\ref{STM1} is a simulation of the STM image of (6,6)@(19,0) computed with
the methodology described in section~\ref{meth}. The tip is at a potential of
0.5~V with respect to the sample. For that polarity, the most prominent
features in the STM image of the semiconducting (19,0) layer are the CC bonds
not parallel to the axis, which appear as bright stripes at 60$^{\circ}$
to the axis.\cite{Kane99} All along the portion of nanotube
displayed, the periodicity of the image is that ($\sqrt{3}a$, with $a$ the
lattice parameter of graphene) of the external zig-zag layer. There is no
visible sign of the underlying armchair nanotube with its shorter period
$a$. The topographic line cut shown at the bottom of fig.~\ref{STM1} clearly
proves that statement. The sharp minima of the curve correspond to the centers
of the hexagons. The apparent variations of their depth are due to the pixel
discretization. The maxima correspond to the atoms, the secondary minima are
at the center of the CC bonds parallel to the axis.

Nothing similar to a Moir\'e pattern appears in the computed STM image of
fig.~\ref{STM1} nor in the simulations we carried out for other polychiral
nanotubes. However, these patterns occur occasionally in the experimental
images of multiwall nanotubes, as reported above. It is not impossible that a
mechanical deformation of outer layer of the tube caused by the STM tip
induces metallic islands (with a much larger density of states) at special
places where the layers are in suitable registry. The pressure of the tip may
also induce better electric contact with the substrate at some places, leading
to a larger tunneling current.

The atoms in fig.~\ref{STM1} all look the same, unlike the case of multilayer
graphite where the STM current at low bias ($\sim$ 0.1~V) shows a strong site
asymmetry.\cite{Tomanek88} This asymmetry is already present with two layers
only,\cite{My185} since the coupling makes the atoms having a neighbor
underneath different from those that have not (A and B atoms, respectively).

In a multiwall nanotube, it is impossible to realize the same stacking as in
natural graphite all around the cross section. However, at least one multiwall
nanotube exists where the topographic STM image is predicted to look like that
of graphite. This case is (5,5)@(10,10).\cite{My185} This system is known to
exhibit small pseudo-gaps near $E_F$ as the consequence of avoided band
crossings~\cite{Kwon98} for relative tube orientations such that the mirror
planes of (5,5) do not coincide with those of (10,10).\cite{My113} Local DOS
calculations then show that the atoms (of the external layer) are not
equivalent, at least in a small interval around the Fermi
level. Interestingly, first-neighbor atoms have a peak or a deep at $E_F$,
alternatively, very much like in graphite where the A and B atoms
alternate. This effect is shown in fig.~\ref{tub10}. The explanation of this
bi-partition of the honeycomb lattice is presented in the Appendix. A
consequence of it is that the STM image of the (5,5)@(10,10) nanotube at low
bias resembles that of graphite, with maxima of protrusion on every other two
atoms (those with the largest DOS at $E_F$), see fig.~\ref{STM2}. However, for
other relative orientations of the tubes, no bipartition effect is observed
(in agreement with first-principle calculations \cite{Rub99,My185}). What is
special about (5,5)@(10,10) is that this system has at least a five-fold
common symmetry. In (6,6)@(11,11) for instance, there is no axial symmetry,
which destroys the effects of the interlayer coupling on the DOS around the
Fermi level, very much like in polychiral nanotubes. The resulting intertube
interaction averaging reduces any symmetry related feature such as the opening
of pseudo-gaps and the bipartition of the honeycomb lattice. The STM image of
(6,6)@(11,11) is then similar to that of the isolated single-wall (11,11)
nanotube (see fig.~\ref{STM2}). With other metallic nanotubes such as
(7,4)@(12,9) which we also have examined, some variations of the local DOS
from one atom to the other were detected in the metallic plateau, but these
were two small to lead to a clear site asymmetry in the STM image.

The property that (5,5)@(10,10) presents a site asymmetry that depends on the
relative orientation of the layers can be illustrated by giving a uniform
torsion to the (10,10) nanotube. At regular intervals along the axis, the
planes that bisect the CC bonds perpendicular to the axis of (10,10) coincide
with the mirror planes of (5,5) ($C_{5v}$ symmetry, no site asymmetry). Away
from these positions, the local symmetry of the atomic structure is lower and
the two-site asymmetry of the DOS should come out and reach a maximum in
between.

The twist is equivalent to applying a shear of the honeycomb network, which
affects the bond lengths and opens a small gap at the Fermi
level.\cite{Rochefort99,Yang99} The calculations were performed for a twist of
1.5$^{\circ}$ (shear strain) corresponding to a torsion angle
of 2.2$^{\circ}$/nm, which leads to a band gap of 0.25~eV (the $\gamma_0$
parameter was scaled according to a $d^{-2}$ law, with $d$ the bond
length). In the (5,5)@(10,10) distorted nanotube, the local DOS of the twisted
(10,10) layer has a peak near the Fermi energy induced by the interactions
with the inner (5,5) nanotube (fig.~\ref{twist}(b)). The shear also affects
the Brillouin zone of the rolled-up graphene sheet, which moves the Fermi
points of the twisted (10,10) nanotube away from those of the (5,5) layer. As
a consequence, the minigaps of the bilayer are no longer located at the center
of the metallic plateau (where the semiconducting gap of the twisted nanotube
has opened) but are shifted 0.25~eV on both sides of the Fermi level. As can
be seen in fig.~\ref{twist}(a), the DOS features in the mini-gaps at $\pm
0.25$~eV resemble the two-site asymmetry observed near $E_F$ in
fig.~\ref{tub10} for the perfect (5,5)@(10,10), except that the magnitude of
the asymmetry now varies along the tube. The DOS curves in fig.~\ref{twist}(a)
correspond to 25 successive atoms along a longitudinal zig-zag chain on the
external tube. The curves at the center look all the same, where the local
symmetry of the (5,5)@(10,10) distorted nanotube is close to $C_{5v}$. By
contrast, the curves at the bottom and at the top have peaks and deeps at $E =
\pm$0.25~eV that alternate from one site to the next. The local symmetry has
been changed to $C_5$ in these regions. The modulation of the degree of
unequivalence between the atoms is due to the continuous change of the local
symmetry along the nanotube axis. Unfortunately, this modulation did not
appear clearly in the STM images that we computed for bias potentials of
$-0.3$ and $+0.3$~V. The reason is that the expression of the STM current
(sect.~\ref{meth}) is an integral of the Green's function elements over the
bias window, to which the site-dependent features in the density of states in
fig.~\ref{twist}(a) contribute little. However, these features might be
observed by current imaging tunneling spectroscopy.

\section{Conclusion}
Two-wall nanotubes mixing metallic and semiconducting layers retain the basic
properties of the uncoupled constituents, as shown previously for monochiral
nanotubes.\cite{Saito93} The intertube interactions induce a small continuous
distribution of states in the band gap of the semiconducting layer. The
electronic states near the Fermi level come from the metallic tube and they
will dominate the transport properties in the weak localization regime.
Constant-current STM images computed for polychiral nanotubes are pretty much
the same as the ones obtained on the isolated external layer. It is only in
the case of monochiral and commensurate structures like (5,5)@(10,10) that
interlayer effects can be seen in the STM topography. The interlayer coupling
gives rise to a site asymmetry in the STM image at low voltage ($\sim$0.1~V),
similar to that obtained on multilayer graphite. This site bipartition is
maximum when the symmetry of the two-wall nanotube is reduced to $C_5$, and it
disappears when the symmetry is higher. The site asymmetry also disappears in
relationally-incommensurate nanotubes like (6,6)@(11,11). In polychiral
nanotubes, there is no site asymmetry and no Moir\'e patterns appear
in the computed STM images. From the few cases we have investigated, it can be
concluded that the superstructures often observed in the STM images of MWNTs
cannot be ascribed to pure electronic effects.

\acknowledgments This work has been partly funded by the interuniversity
research project on reduced dimensionality systems (PAI P4/10) of the Belgian
Office for Scientific, Cultural and Technical affairs, the EU NAMITECH
contract: ERBFMRX-CT96-0067 (DG12-MITH) and by a JCyL (Grant: VA28/99).  The
authors acknowledge Laszlo P. Bir\'o for helpful discussions.

\section*{Appendix: Multi-wall armchair nanotubes} 
In a single-wall armchair nanotube, the two bands that cross each-other at the
Fermi level correspond to the irreducible representation $A_1$ and $A_2$ of
the symmetry group $C_{5v}$ of the wave function for a general Bloch wave
vector $k$. These wave functions are respectively symmetric and antisymmetric
upon a reflection on the ``vertical'' mirror planes that bisect the CC bonds
perpendicular to the nanotube axis. When all curvature effects are neglected
in the Hamiltonian as here, the Fermi wave vector is independent on the tube
diameter. This means that in a two-wall nanotube such as (5,5)@(10,10), the
states at the Fermi points have a fourfold degeneracy when the interlayer
coupling is ignored. In the presence of the coupling, the wave functions
adapted to the perturbation $W$ are linear combinations of the four Fermi
states $\psi_1^i$, $\psi_2^i$, $\psi_1^e$, and $\psi_2^e$ of the internal and
external layers (upper indices $i$ and $e$) corresponding to $A_1$ and $A_2$
symmetries (lower indices 1 and 2). These combinations of states diagonalize
the perturbation matrix
\begin{displaymath}
\left( \begin{array}{llll}
	0 & 0 & w_{11}^{ie} & w_{12}^{ie} \\
	0 & 0 & w_{21}^{ie} & w_{22}^{ie} \\
	w_{11}^{ei} & w_{12}^{ei} & 0 & 0 \\
	w_{21}^{ei} & w_{22}^{ei} & 0 & 0 
       \end{array}
\right)
\end{displaymath}
where $w_{12}^{ie} = \langle \psi_1^i | W | \psi_2^e \rangle$, etc. In any
case, the perturbation is sufficient to split off the degeneracy of the Fermi
states. In general, also, the eigenvector of the perturbation matrix will mix
the four Fermi states. This means in particular that the perturbed wave
functions at the Fermi level mix the $\psi_1^e$ and $\psi_2^e$ states of the
external layer. By mixing these states, which are respectively even and odd
with respect to the center of each CC bond, one forms a disymmetric
combination. As a result, there are now two kinds of unequivalent atoms in
each layer, as shown by the density of states in fig.~\ref{tub10}, which
explains the site asymmetry of the computed STM images.

An exception to this explanation arises when the four matrix elements of the
kind $w_{12}^{ie}$ along the ascending diagonal vanish for symmetry
reason. This takes place when the symmetry of the two-wall nanotube preserves
the mirror planes of the inner layer. Then, $A_1$ and $A_2$ remain valid
irreducible representations of the symmetry group of the coupled system, band
crossings remain allowed, and there is no pseudo-gap formation in the density
of states. This also means that the site asymmetry of the STM image
disappears. For the (5,5)@(10,10) nanotube, this happens with $C_{5v}$,
$D_{5h}$ and $D_{5d}$ configurations of the layers.\cite{My113,My126,Kwon98}
The latter configuration corresponds to the minimum of the total energy of the
nanotube.\cite{Charlier93,Kwon98}

In the case where the two nanotubes have no axial symmetry in common, such as
for instance with (6,6)@(11,11), all band crossings are avoided and two kinds
of unequivalent atoms are formed in each layer, as above. However, due to a
cancellation effect, all the elements in the perturbation matrix are found to
be small except the ones derived from the totally-symmetric states,
$w_{11}^{ie}$ and $w_{11}^{ei}$. In practice, then, all the atoms look
equivalent. Also, the pseudo-gaps near the Fermi level are much weaker
than for the (5,5)@(10,10) case.\cite{My126}

\begin{table}[h]
\caption{Two-wall nanotubes with a metallic inner shell and a semiconducting
outer shell. The second row is the intersheet distance, the third is the
number of states induced in the gap, and the fourth column is the standard
deviation of the local density of states in the external layer (see
sect.~\ref{STMsec}).}
\begin{tabular}{cccc}
nanotube &$R_2-R_1$ (nm) &gap states/atom &rms (eV$^{-1}$/atom)\\ \hline
(6,6)@(19,0) &0.337 &0.17$\times10^{-4}$ &3.7$\times10^{-4}$\\
(6,6)@(18,2) &0.340 &0.28$\times10^{-4}$ &2.4$\times10^{-4}$\\
(10,-2)@(16,3) &0.337 &0.56$\times10^{-4}$ &2.7$\times10^{-4}$\\
(15,-6)@(15,10) &0.341 &0.42$\times10^{-4}$ &3.2$\times10^{-4}$\\
(9,6)@(15,10) &0.341 &2.28$\times10^{-4}$ &2.8$\times10^{-4}$\\
\end{tabular}
\label{list1}
\end{table}

\begin{table}[h]
\caption{Same as table~\ref{list1} for polychiral two-wall nanotubes having a
semiconducting inner shell and a metallic outer shell.}
\begin{tabular}{cccc}
nanotube &$R_2-R_1$ (nm) &gap states/atom &rms (eV$^{-1}$/atom)\\ \hline
(6,4)@(10,10) &0.337 &2.4$\times10^{-4}$ &7.6$\times10^{-4}$ \\
(7,3)@(13,7) &0.340 &2.2$\times10^{-4}$ &5.0$\times10^{-4}$ \\
(7,6)@(13,10) &0.341 &2.1$\times10^{-4}$ &5.5$\times10^{-4}$ \\
(10,0)@(15,6) &0.342 &2.6$\times10^{-4}$ &6.1$\times10^{-4}$ \\
\end{tabular}
\label{list2}
\end{table}

\begin{figure}[h]
\caption{Average density of states (eV$^{-1}$/atom) in the external layer of
(a) (15,-6)@(15,10) and (b) (9,6)@(15,10) nanotubes (full curves). The dashed
curve is the density of states of the single-wall (15,10) nanotube. The
differences between the DOS's of the two-layer and one-layer systems are shown
by a dotted curve, displaced vertically for clarity (the horizontal bar
indicates zero difference), and multiplied by a factor of
10.}
\label{tub4/5}
\end{figure}

\begin{figure}[h]
\caption{Average density of states in (a) the inner and (b) the outer layer of
(6,4)@(10,10) (full curves). The dashed curves in (a) and (b) show the density
of states of the isolated (6,4) and (10,10) nanotubes. The differences between
the full and dashed curves are shown by the dotted curves after amplification
by a factor of 10.}
\label{tub6}
\end{figure}

\begin{figure}[h]
\caption{Top: computed constant-current image of the topmost part of the
(6,6)@(19,0) two-layer nanotube. The nanotube axis is parallel to the
horizontal $x$ direction. The tip potential is 0.5~V, the height of the tip
above the atom at the origin of the $x,y$ coordinates is 5~\AA\ (all the
coordinates are in \AA). Bottom: topographic line cut across the image, along
the $y=0$ line.}
\label{STM1}
\end{figure}

\begin{figure}[h]
\caption{$\pi$-electron band structure of (5,5)@(10,10) and local density of
states on the sites labeled 1-8 of the external layer. The configuration of
the nanotube, with symmetry $C_{5h}$, is sketched in the top-left part, with a
cross section (large and small dots correspond to the atoms located in the
plane of the drawing and those outside the plane, respectively), and the
planar development of a short azimuthal portion of the external layer (with
the 8 labeled sites, the projections of the inner sites being indicated by
black dots).}
\label{tub10}
\end{figure}

\begin{figure}[h]
\caption{Gray-scale representation of the radial distance $\rho(x,y)$ at
constant current of the STM tip apex above a short azimuthal portion of
(5,5)@(10,10) (left) and (6,6)@(11,11) (right). The orientation of the two
layers in (5,5)@(10,10) is the $C_{5h}$ configuration of fig.~\ref{tub10}. The
tube axes are along the horizontal direction and the tip potential is fixed at
0.1 V.}
\label{STM2}
\end{figure}

\begin{figure}[h]
\caption{Local DOS in the outer layer of (5,5)@(10,10) obtained by twisting
the (10,10) nanotube with a uniform torsion equal to 2.2$^{\circ}$/nm. (a)
Variations of the local DOS along a chain of 25 atoms located as close as
possible to a generator of the external tube. (b) The average of the 25 curves
shown in (a) (full curve) is compared to the DOS of the isolated twisted
(10,10) nanotube (dashed curve).}
\label{twist} \end{figure} 
\begin{thebibliography}{99}
\bibitem{Saito93}R. Saito, G. Dresselhaus, and M.S. Dresselhaus,
J. Appl. Phys. {\bf 73}, 494 (1993).
\bibitem{Charlier93}J.C. Charlier and J.P. Michenaud, Phys. Rev. Lett. {\bf
70}, 1858 (1993).
\bibitem{My113}Ph. Lambin, L. Philippe, J.C. Charlier, and J.P. Michenaud,
Comput. Mater. Sci. {\bf 2}, 350 (1994).
\bibitem{My126}Ph. Lambin, J.C. Charlier, and J.P. Michenaud, in {\it Progress
in fullerene research}, edit. H. Kuzmany, J. Fink, M. Mehring, and S. Roth
(World Scientific, Singapore, 1994), 130-134.
\bibitem{Ostling97}D. \"Ostling, D. Tom\'anek, and A. Ros\'en, Phys. Rev. B
{\bf 55}, 13980 (1997).
\bibitem{Kwon98}Y.K. Kwon and D. Tom\'anek, Phys. Rev. B {\bf 58}, R16001
(1998).
\bibitem{Rub99}A. Rubio, Appl. Phys. A {\bf 68} , 275 (1999).
\bibitem{Iijima91}S. Iijima, Nature {\bf 354}, 56 (1991).
\bibitem{Zhang93}X.F. Zhang, X.B. Zhang, G. Van Tendeloo, S. Amelinckx, M. Op
de Beeck, and J. Van Landuyt, J. Crystal Growth {\bf 130}, 368 (1993).
\bibitem{Liu94}M. Liu and J.M. Cowley, Carbon {\bf 32}, 393 (1994).
\bibitem{Iijima94}S. Iijima, MRS Bull. Nov 1994, 43-9.
\bibitem{Ge93}M. Ge and K. Sattler, Science {\bf 260}, 515 (1993).
\bibitem{Hassanien99}A. Hassanien, M. Tokumoto, S. Ohshima, Y. Kuriki,
F. Ikazaki, K. Uchida, and M. Yumara, Appl. Phys. Lett. {\bf 75}, 2755
(1999).
\bibitem{Anderson58}, P.W. Anderson Phys. Rev. {\bf 109}, 1492 (1958).
\bibitem{Lee85}P.A. Lee and T.V. Ramakrishnan, Rev. Mod. Phys. {\bf 57}, 287
(1985) and references therein.
\bibitem{Lin96}N. Lin, J. Ding, S. Yang, and N. Cue, Carbon {\bf 34}, 1295 (1996).
\bibitem{Biro98}L.P. Bir\'o, J. Gyulai, Ph. Lambin, J. B.Nagy, S. Lazarescu,
G.I. M\'ark, A. Fonseca, P.R. Surj\'an, Zs. Szekeres, P.A. Thiry, and
A.A. Lucas, Carbon {\bf 36}, 689 (1998).
\bibitem{Saito00}R. Saito, G. Dresselhaus, and M. S. Dresselhaus, Phys. Rev. B
{\bf 61}, 2981 (2000).
\bibitem{My174}V. Meunier and Ph. Lambin, Phys. Rev. Lett. {\bf 81}, 5888
(1998).
\bibitem{My185}Ph. Lambin, V. Meunier, and A. Rubio, in {\it Science and
Applications of Carbon Nanotubes}, edit. D. Tom\'anek and R.J. Enbody (Kluwer
Academic Publisher, New York, 2000), 17-34.
\bibitem{Langer96}L. Langer, V. Bayot, E. Grivei, J.P. Issi, J.P. Heremans,
C.H. Olk, L. Stockman, C. Van Haesendonck, and Y. Bruynseraede,
Phys. Rev. Lett. {\bf 76}, 479 (1996).
\bibitem{Schonenberger99}Sch\"onenberger, A. Bachtold, C. Strunk,
J.P. Salvetat, and L. Forr\'o, Appl. Phys. A {\bf 69}, 283 (1999).
\bibitem{Beyer99}H. Beyer, M. M\"uller, and Th. Schimmel, Appl. Phys. A {\bf
68}, 163 (1999).
\bibitem{Kane99}C.L. Kane and E.J. Mele, Phys. Rev. B {\bf 59}, R12759 (1999).
\bibitem{Tomanek88}D. Tom\'anek and S.G. Louie, Phys. Rev. B {\bf 37}, 8327
(1988).
\bibitem{Rochefort99}A. Rochefort, Ph. Avouris, F. Lesage, D.R. Salahub,
Phys. Rev. B {\bf 60}, 13824 (1999).
\bibitem{Yang99}L. Yang, M.P. Anantram, J. Han, and J.P. Lu, Phys. Rev. B {\bf 60}, 13874 (1999).
\end{thebibliography}
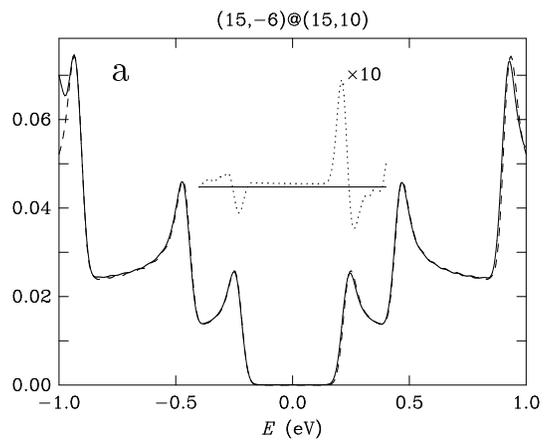
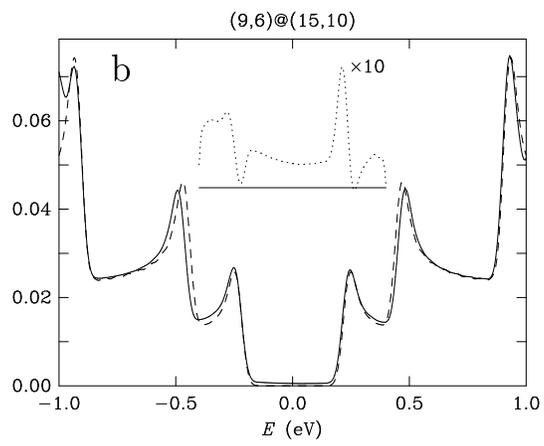
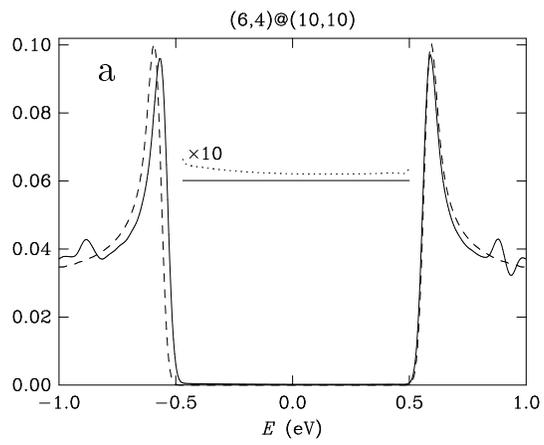
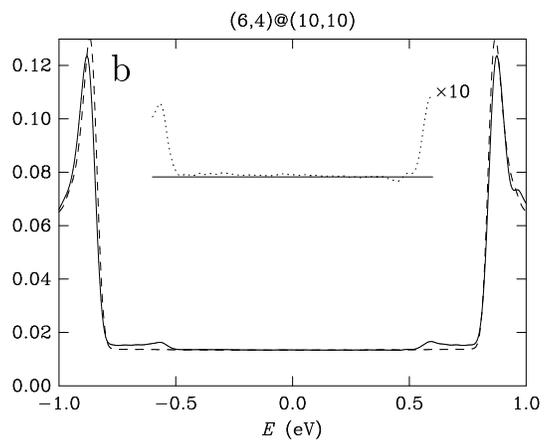
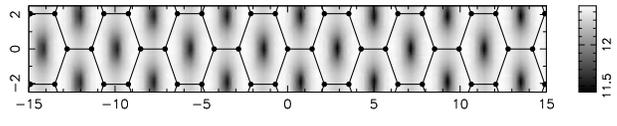
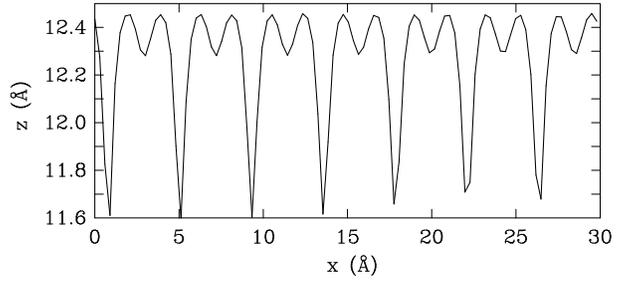
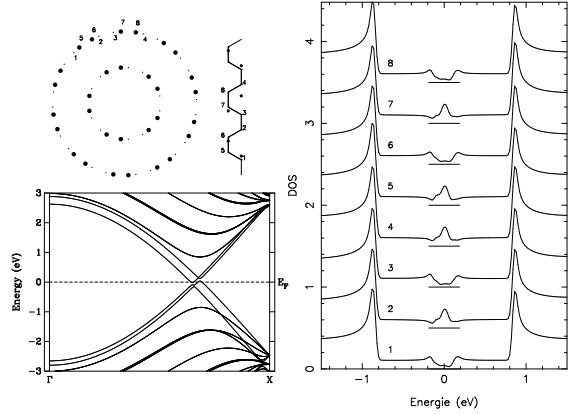
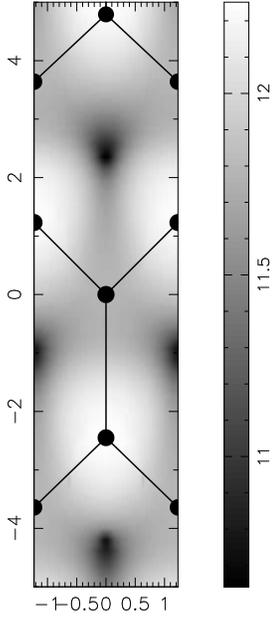
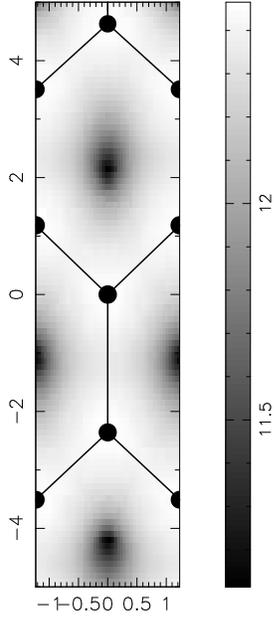
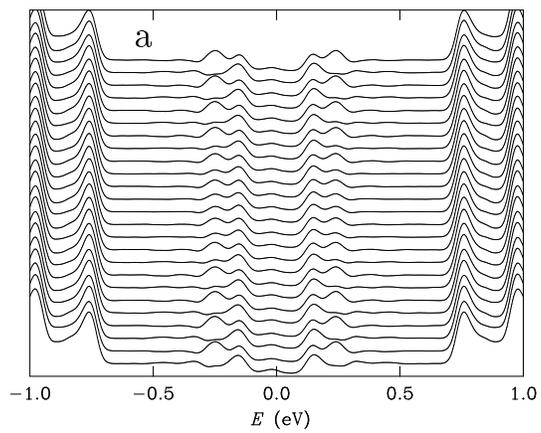
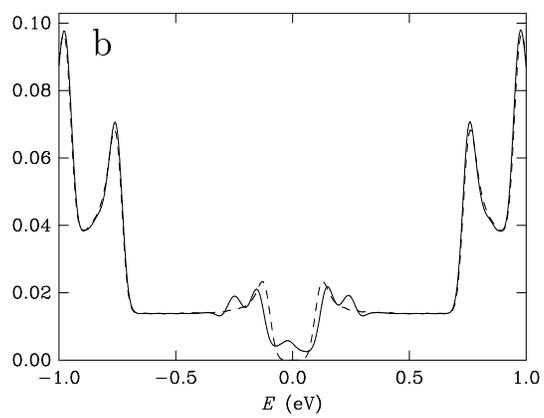
\end{document}